\documentclass{iopart}
\usepackage{graphicx}
\usepackage{cite}
\usepackage{amssymb}
\usepackage{color}
\begin{document}

\title{Electronic and Optical Properties of Electromigrated Molecular Junctions}
\author{D R Ward$^{1}$, G D Scott$^{1}$, Z K Keane$^{1}$, N J Halas$^{2,3}$, D Natelson$^{1,2}$}
\address{$^{1}$ Department of Physics and Astronomy, Rice University, Houston, TX 77005, USA}
\address{$^{2}$ Department of Electrical and Computer Engineering, Rice University, Houston, TX 77005, USA}
\address{$^{3}$ Department of Chemistry, Rice University, Houston, TX 77005, USA}
\eads{\mailto{natelson@rice.edu},~\mailto{daniel.ward@rice.edu},~\mailto{gavin.scott@rice.edu}}

\date{\today}

\begin{abstract}
Electromigrated nanoscale junctions have proven very useful for
studying electronic transport at the single-molecule scale.  However,
confirming that conduction is through precisely the molecule of
interest and not some contaminant or metal nanoparticle has remained a
persistent challenge, typically requiring a statistical analysis of
many devices.  We review how transport mechanisms in both purely
electronic and optical measurements can be used to infer information
about the nanoscale junction configuration.  The electronic response
to optical excitation is particularly revealing.  We briefly
discuss surface-enhanced Raman spectroscopy on such junctions,
and present new results showing that currents due to optical
rectification can provide a means of estimating the local electric
field at the junction due to illumination.

\end{abstract}

\maketitle

\section{Introduction}
Electronic devices with molecules and organic components as active
elements offer new limits of device scaling and functionality, and are
also of fundamental physical interest.  Studies at the single molecule
level probe the physics of electronic conduction and optical
interactions in regimes that have been previously inaccessible.
Three-terminal electronic measurements have proven invaluable in many
systems, enabling electronic transport to function as a spectroscopy
of available states.\cite{Ahn2006} While much progress has been made
using two-terminal measurements to examine charge transport through
molecules\cite{Ruitenbeek1996,Datta1997,Reed1997,Scheer1998,Stipe1998,Chen1999,Collier1999,Donhauser2001,Wold2001,Cui2001,Reicher2002,Smit2002,Kushmerick2002,Ramachandran2003,Xu2003,Xiao2004,Tao2006,Venkataraman2006,Venkataraman2007}
, three-terminal measurements greatly increase the available
information in single-molecule measurements.  For example, through the
use of a gate electrode it is possible to study conduction as a
function of the molecule's redox potential, analogous to what can be
done in electrochemical experiments.

A common approach to fabricating three-terminal single-molecule
devices is known as the ``electromigration technique'', in which
thermally assisted electromigration is used to create a nanoscale gap
in a narrow metallic wire situated on a gate insulator
material\cite{HPark1999}.  If a molecule of interest resides in the
nanogap, the broken ends of the wires are used as the source and drain
contact electrodes, and an underlying conductive substrate can be used
as a gate electrode.  The resulting device is a single-molecule
transistor (SMT).  Electromigrated breakjunctions have been used to
study transport through individual nanocrystals\cite{HPark1999}, small metal particles\cite{Bolotin2004}, a variety of
molecules\cite{HPark2000,JPark2002,Liang2002,Yu2004b,Pasupathy2004,Yu2004c,Pasupathy2005,Champagne2005,Yu2005,Heersche2006,Chae2006,Danilov2006,vanderZant2006,Scott2006,Natelson2006,Osorio2007,Danilov2008},
and even individual atoms\cite{Heersche2006b}.

A central hurdle in most single-molecule electronic measurements is to
demonstrate unambiguously that transport is occurring through only a
single molecule of interest.  This is complicated by the lack of
direct imaging techniques with sufficient resolution, except for the
two-terminal example of scanning tunneling microscopy (STM), which
allows imaging with atomic resolution on conducting substrates.  In
this paper we review SMT fabrication (Section~\ref{fab}), and then
discuss device characterization based on electronic transport alone as
a diagnostic (Section~\ref{transport}).  In that circumstance it is
necessary to search for features characteristic of conduction through
a single molecule in addition to attributes uniquely identifying the
molecule of interest.  Generally the search for single molecule
devices requires a statistical approach and many
control experiments.  Exciting recent work (Section~\ref{optical})
demonstrates that simultaneous single-molecule optical spectroscopy
and transport is possible.  In such measurements surface-enhanced
Raman spectroscopy (SERS) can give {\it the vibrational spectrum of
the specific molecule through which transport is taking place}.  We
present new data using nonlinear transport and optical rectification
to arrive at a quantitative estimate for the local optical field
experienced by the molecule under illumination.

\section{Device Fabrication}
\label{fab}

\begin{figure}[!h]
\begin{center}
\includegraphics [clip, width=8 cm]{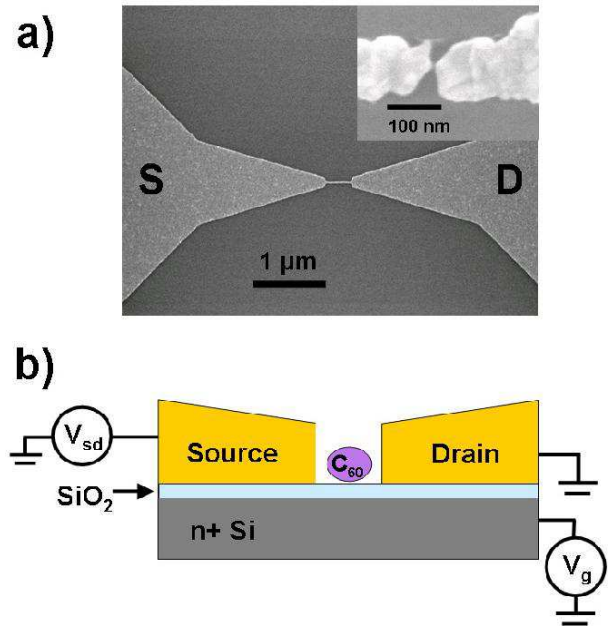}
\end{center}
\hspace\fill \vspace{-8mm} \caption{(a) SEM image of a gold nanowire structure on Si/SiO$_2$ prior to electromigration.  Inset shows a magnified SEM image of a similar nanowire after the breaking procedure.  (b) Ideal device configuration after the electromigration process is complete for a C$_{60}$-based junction.} \label{SEMImage}
\end{figure}

Fabrication begins with the preparation of arrays of
devices on an $n^+$ Si substrate with a 200 nm SiO$_2$ insulating
layer.  1 nm of Ti and 15 nm of Au are evaporated onto nanowire
patterns defined by electron beam lithography.  Bowtie-shaped
constriction patterns are produced with minimum widths of
approximately 100~nm (Figure \ref{SEMImage}a).  Additional gold pads
for contacting the source and drain electrodes are also defined by
electron beam lithography.  The array of samples is cleaned in
solvents then exposed to an oxygen plasma for 1 min to remove trace
organics.   The molecules are dispersed for 
approximately monolayer coverage, or allowed to self-assemble
onto the Au surface, depending on the molecule in question.

Electromigration (with some thermal enhancement due to Joule heating)
is used to break the constriction into distinct source and drain
electrodes, with the intent that the molecule will reside in the
resulting nanoscale gap.  This process has been studied extensively by
a number of
investigators\cite{Strachan2005,Trouwborst2006,Strachan2006,Dong2006,Ralph2007,Oneill2007}.
At large current densities in the constriction, the electrons transfer
sufficient momentum to the lattice to move atoms at the surface and at
grain boundaries.  As a constriction is formed the local current
density increases, leading to a runaway migration and the formation of
an interelectrode gap.  Elevated local temperatures enhance the rate
of this process through the increased diffusion of metal atoms.  To
mitigate concerns about adsorbed contaminants we prefer to
electromigrate junctions at low temperatures in a cryopumped ultrahigh
vacuum (UHV) environment, though this has not been possible in the
optical experiments discussed in Section~\ref{optical}.  The critical
gap size, defined as the minimum separation between the two resulting
electrodes, is typically 1-3~nm.  With good, reproducible lithography
to produce the constrictions and an appropriate electromigration
procedure, yields of such small gaps can exceed 90\%.

An assortment of voltage ramping techniques have been utilized by
different research groups in an effort to form the smallest, cleanest,
and most consistently-sized
nanogaps\cite{Strachan2005,Trouwborst2006,Strachan2006,Dong2006,Ralph2007,Oneill2007}.
All techniques involve minimizing the series resistance to avoid
overheating and subsequent melting of a nanowire; and an immediate
reduction in applied bias across the break after the gap has opened in
order to avoid creating an excessively large interelectrode spacing.
We have found success with a feedback controlled method of repeatedly
ramping the source-drain voltage, $V_{\mathrm{SD}}$.

The desired final device configuration is shown schematically in
Figure \ref{SEMImage}b.  Subtle differences in junction configuration
down to the {\AA}ngstr{\"o}m level can strongly influence device
properties.  The WKB approximation dictates that the current through a
metal-vacuum-metal gap decreases by roughly a factor of $e^2$
($\sim$7.4) for every {\AA}ngstr{\"o}m increase in the size of the
vacuum gap\cite{Wolf1985}.  This exponential dependence of tunneling
on interelectrode distance ensures, as in an STM, that the tunneling
electrons probe a volume containing at most one or two molecules.
Precisely how molecules that begin on the surface of the constriction
end up in the interelectrode gap is not known.  The inherent
randomness in the device formation process leads to many
electrode-molecule configurations and resulting electrical conduction
characteristics for any group of devices.  Historically this has
mandated a statistical approach to device characterization based on
transport.  By analysis of the transport data, as explained below, it
is often possible to infer the number of molecules in the tunneling
region and the relative couplings of the molecule to the source and
drain electrodes.  Details about molecular positioning and bonding
remains unavailable, though new measurements such as those in
Sect.~\ref{optical} contain much information.

\section{Low temperature electronic transport}
\label{transport}
\begin{figure}[!h]
\begin{center}
\includegraphics[clip, width=8cm]{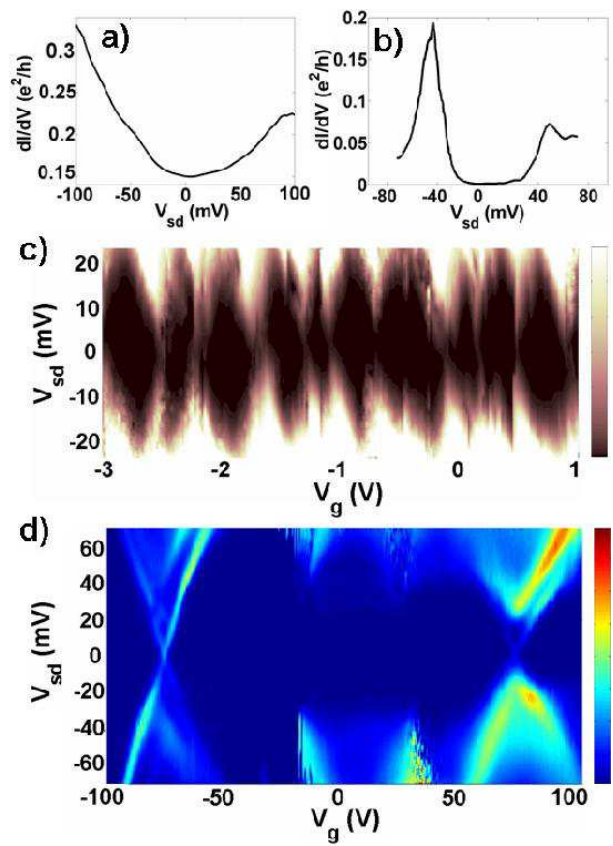}
\end{center}
\hspace\fill \vspace{-8mm} \caption{(a) Plot of $dI/dV$ vs $V_{\mathrm{SD}}$ demonstrating a device exhibiting conduction in the ``weakly nonlinear'' regime. (b) Plot of $dI/dV$ vs $V_{\mathrm{SD}}$ for a device exhibiting a Coulomb blockade.  (c)  Stability diagram of a device with clear gate dependence.  Black represent zero conductance and white represents 0.003 $e^2/h$.  From equation 1, the charging energy is approximately 20 meV.  The small charging energy along with the high number of accessible charge states indicates that the active element is not a single molecule of interest, but rather a metal particle.  (d) Stability diagram of device that possesses a single C$_{60}$ molecule as the active element.  $E_{\mathrm{c}} > 300$~meV.  Dark blue represents zero conductance and red represents 0.4 $e^2/h$.  The visible excited states are consistent with vibrational excitations of the molecule.} \label{Blockades2}
\end{figure}

The differential conductance, $dI/dV$, as a function
of $V_{\mathrm{SD}}$, calculated or measured using quasi-dc techniques,
is the main transport tool to assess the nature of a given device.  
There are four main
classes of differential conductance traces, with characteristic
abundances determined by examining large numbers (thousands) of
junctions\cite{Natelson2006} (though these vary depending on molecule
type).  1) Devices with no measurable differential conductance
indicate an interelectrode breakjunction gap that is too large for
measurable conduction by tunneling ($\sim 10$\%).  2) Devices with a
weakly nonlinear conductance plot wherein $dI/dV$ drops around
$V_{\mathrm{SD}}$ = 0, but does not form a clear blockaded region
denote breakjunctions with small to moderately sized gaps, but without
an active element within the nanogap (Figure \ref{Blockades2}a, $\sim
45$\%).  3) Devices that show a region of zero
conductance bordered by sharp peaks most likely have a 
gap containing a molecule or small particle (Figure
\ref{Blockades2}b, percentage depends strongly on molecule type).  
4) Devices with a zero-bias resonance suggest a stronger
molecule-electrode coupling (percentage depends strongly on
molecule type, ranging from zero for alkane chains to 
several percent for transition metal complexes containing
unpaired $d$ electrons\cite{Yu2005}).  

Further investigation is required to identify those devices that
contain a single molecule of interest.  Confounding possibilities
include surface contaminants, metal grains produced during the
electromigration process, or simply more than one molecule of
interest.  As explained below, the observed gate dependence of
conduction and an understanding of the relevant transport physics has
enabled progress in eliminating spurious devices.  The final yield of
gateable likely SMTs is typically 10-15\% of the starting devices,
based on a total sample size of thousands of devices.

\subsection{Coulomb blockade}
In analogy to conventional single-electron transistors, the SMT can be
thought of as consisting of 6 parts: source, drain and gate
electrodes, the molecule, and the tunneling connections to the source
and drain.  The tunneling barriers are established by the geometry and
chemistry of the molecule/electrode interface.  We assume that the
polarizability of the complex is much larger than that of the
insulating barriers; therefore the voltage drops on the barriers are
proportional to their respective coupling strengths ${\Gamma}_{\mathrm{S}}$ and
${\Gamma}_{\mathrm{D}}$.  Elementary transport characteristics can be calculated
by treating tunneling barriers as a capacitive and
a resistive element\cite{Ferry1997}. The positive and negative slopes
of the blockaded region (the tunneling thresholds) are determined by
the ratios $C_{\mathrm G}/C_{\mathrm{D}}$ and $C_{\mathrm G}/C_{\mathrm{S}}$,
respectively\cite{Champagne2005} ($C_{\mathrm G}$, $C_{\mathrm{D}}$, and $C_{\mathrm{S}}$ are the
capacitances between the molecule-gate, molecule-drain, and
molecule-source electrodes).

Traces of $dI/dV$ measured as a function of $V_{\mathrm{SD}}$,
obtained within a range of incrementally changing gate voltage,
$V_{\mathrm{G}}$, can be compiled to form a map known as a stability
diagram, illustrating where tunneling is both allowed and prohibited.
Not all devices exhibiting a zero-bias resonance or a conductance gap
will show dependence on applied gate bias, $V_{\mathrm G}$.  This poor
gate coupling may occur as a result of the detailed device geometry or
possibly as a result of the orientation of the active element within
the breakjunction gap\cite{Yu2005,Perrine2007}. Stability diagrams of
devices that \emph{do} exhibit gate dependence are a primary means of
displaying pertinent transport data reflecting important aspects of
SMT behavior.

\begin{figure}[!t]
\begin{center}
\includegraphics [clip, width=8cm]{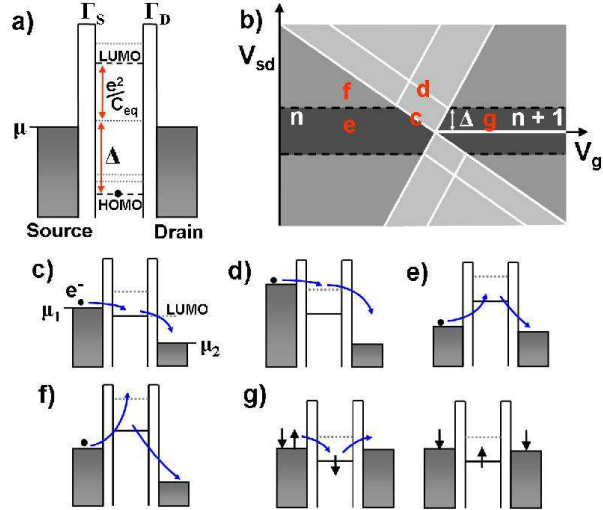}
\end{center}
\hspace\fill \vspace{-8mm} \caption{(a) Energy level diagram of a single molecule transistor.  $\Delta$ is the single particle level spacing.  (b) Stability diagram of a model SMT displaying dI/d$V_{\mathrm{SD}}$ (shown as relative brightness) as a function of $V_{\mathrm{SD}}$ and $V_{\mathrm G}$.  A transition from the charge state $n$ to the state $n + 1$ is shown for a molecule with $n$ electrons.  (c) Resonant tunneling.  (d) Inelastic resonant tunneling. (e) Elastic cotunneling.  (f)  Inelastic cotunneling.  (g) Cotunneling leading to the Kondo effect.} \label{Tunneling1}
\end{figure}

In equilibrium the source and drain electrodes have a common
electrochemical potential, $\mu$, the Fermi level.  When the source
and drain electrodes are biased such that the source electrode has a
chemical potential $\mu_1$ and the drain electrode has chemical
potential $\mu_2$, current will flow as long as the Coulomb charging
energy ($E_{\mathrm{c}}$) and the discrete energy level spacing
($\Delta$) have been overcome, and an energy level of the molecule
lies between $\mu_1$ and $\mu_2$ (at finite temp, this range is
extended by the thermal energy $\pm k_{\mathrm B}T$\cite{Ghosh2002}).  This
current flow is due to resonant tunneling of electrons from the source
to the lowest available single particle state producing a $dI/dV$ peak
at the edge of the conductance gap (Figure \ref{Tunneling1}c).
Gate voltage shifts the active element levels
with respect to $\mu_1$ and $\mu_2$.  The
resulting stability diagram is determined by both the active element's
spectrum and its capacitive couplings to the source, drain, and
gate electrodes.

When the active element has a negligible single-particle level spacing
(\emph{e.g.}, a few-nm metal grain), the discrete spectrum seen in
transport arises purely from $E_{\mathrm{c}}$, the Coulomb repulsion
that must be overcome to add another electron to the element.  The
result is a series of regularly spaced diamonds in the stability
diagram, and the suppressed conduction in the absence of a discrete
level available for transport is called Coulomb blockade.  Each
diamond region represents a different charge state of the particle
(\emph{i.e.} a stable region of fixed average electron number).  In
this case the electron addition energy depends upon the classical
capacitance of the particle, including interactions with the
electrodes. 

Frequently the single-particle electronic level spacing, $\Delta$,
cannot be neglected when considering transport.  If the island is a
molecule such as C$_{60}$, the discrete energy level spacing between
single particle energy states at equilibrium occurs between the
highest occupied molecular orbital level (HOMO level) and the lowest
unoccupied molecule orbital level (LUMO level), commonly called the
HOMO-LUMO gap.  We can see from equation (1) that the size of the
Coulomb diamond centered around $V_{\mathrm{SD}},~V_{\mathrm G} = 0$
depends on both the Coulomb charging energy as well as the magnitude
of the HOMO-LUMO gap.
\begin{equation}
E_{\mathrm{c}} = e\frac{C_{\mathrm{G}}}{C_{\mathrm{eq}}}\Delta{V_{\mathrm G}} = \frac{1}{2}e\left(\frac{dV_{\mathrm{SD}}}{dV_{\mathrm G}}\right)\Delta{V_{\mathrm G}},
\end{equation}
where $C_{\mathrm{eq}} = C_{\mathrm G} + C_{\mathrm{S}} + C_{\mathrm{D}}$, $dV_{\mathrm {SD}}/dV_{\mathrm G}$ is the slope of the diamonds, and $\Delta V_{\mathrm G}$ is the
voltage spacing between successive charge states.
\begin{equation}
\Delta V_{\mathrm G} = \frac{C_{\mathrm{eq}}}{eC_{\mathrm G}(\Delta + \frac{e^2}{C_{\mathrm{eq}}})},
\end{equation}
where $\Delta$ is the spacing between two discrete energy levels
(e.g. the HOMO-LUMO gap), and $e^2/C_{\mathrm{eq}}$ is the bare Coulomb
charging energy, $E_{\mathrm{c}}$.

Unlike metal particles, most molecules have a very limited number of
accessible charge states, which do not typically appear with even
spacing.  Solution-based electrochemistry (e.g. cyclic voltammetry)
demonstrates that molecules have a limited number of accessible
valence states.\cite{Natelson2006} If one finds that it is possible to
add many electrons to a potential device via gating, or if the spacing
between charge state transitions is highly regular, it is extremely
unlikely that the active region of the potential device contains a
single small molecule.  

Stability diagrams with linear tunneling barrier edges possessing only
two characteristic slopes (one positive and one negative) can be
simulated using the basic capacitance model of sequential electron
tunneling\cite{Scott2006}.  A system with more that one active element
will always have more than two characteristic slopes.  This provides a
means of assessing whether tunneling is occurring through multiple
molecules.

When scrutinizing a potential SMT with a clear blockaded region, one
must also look at the electron addition energy (the max source-drain
bias of the blockaded region), which should be of a sensible size for
the molecule in question.  The classical capacitance of a molecule is
so small that an electrostatic charging energy over 100~meV is not
unreasonable, even without taking into account molecular level
spacing.  

%\section{Detailed Measurements}

Some characteristics in stability diagrams may uniquely identify an
SMT made with a particular molecule.  At sufficiently high source
drain bias, an electron may tunnel from the source into an
\emph{excited} single particle state of the island (Figure
\ref{Tunneling1}d).  Additional $dI/dV$ peaks that parallel the edges
of the blockaded diamonds in stability diagrams appear beyond the edge
of the conductance gap when tunneling occurs through newly accessible quantized excitations.  The location of each $dI/dV$ peak outside
the conductance gap provides information on the excitation spectrum 
of a SMT.

The conductance gap disappears (ordinarily) at the charge degeneracy
point, $V_{\mathrm G} = V_{\mathrm{c}}$, where the total energy of the
system is the same for two different charge states of the molecule.
When the gate voltage passes V$_{\mathrm{c}}$ as it is ramped in the positive
direction, the average number of charges on the molecule changes by
one electron.  The absolute equilibrium electron population, $n$, is
determined by $V_{\mathrm G}$, molecule/electrode interfacial charge
transfer, and the local charge distribution around the molecule.
Unfortunately, $n$ cannot be determined solely from the data
obtained in the stability diagram.

Each $dI/dV$ peak on the $V_{\mathrm G} < V_{\mathrm{c}}$ side of a
stability diagram represents an opening of a new tunneling pathway
where an electron tunnels onto the $n$-electron state, for example, to
transiently generate the $(n+1)$-electron ground or excited states.
In this way, the peaks probe the excitation energies of the molecule.
Conversely, each $dI/dV$ peak that appears at $V_{\mathrm G}>
V_{\mathrm{c}}$ probes the ground and excited states of the
$n$-electron molecule\cite{JPark2002}.  The energy of these quantized
excitations can be determined from the source/drain voltage at which
they intercept the conductance gap.  \emph{If these excitations can be
  identified with known molecular properties, they can serve as a
  fingerprint for molecular identification.}

Higher-order tunneling events known as cotunneling may be observed
inside the Coulomb blockade region of an SMT when the overall
tunneling current is very small.  Cotunneling becomes more apparent as
the coupling between the dot and leads is enhanced, and can give rise
to the non-zero current inside the blockaded region.  Elastic
cotunneling corresponds to an electron tunneling into and out of the
same energy level, such that the molecule remains in its ground state
(Fig \ref{Tunneling1}e).  Inelastic cotunneling occurs when
an electron enters and exits the molecule through two different energy
levels, ultimately leaving the molecule in an excited state.  When
$eV_{\mathrm{SD}}= \delta$, where $\delta$ is the energy level spacing
between the ground state and the first excited state, inelastic
cotunneling processes can occur (Figure~
\ref{Tunneling1}f)\cite{Franceschi2001,Sukhorukov2001}.  Cotunneling
events are only weakly affected by $V_{\mathrm G}$ and can be most
clearly seen in a stability diagram mapping $d^2I/dV^2$.  Vibrational
inelastic cotunneling is equivalent to inelastic electron tunneling
spectroscopy\cite{Yu2004c}.

Excited states can originate from several possible degrees of freedom,
including excited electronic states of the system and, in molecules,
internal vibrational modes.  If the pattern of excitations is
identical for multiple charge states of a given molecule
(i.e. excitation are observed on both sides of $V_{\mathrm{c}}$ at the
same value of $V_{\mathrm{SD}}$) this suggests that the excitations
are not associated with molecular charge states and are therefore
independent of the electronic configuration.  Additionally, if
excitation energies are equally spaced then their origin is also
unlikely to be excited electronic charge states.

As long as the exact structure of a molecule is known, its vibrational
modes can be computed, and further calculations can demonstrate that
only some of these modes are plausible candidates for excitation peaks
observed in stability diagrams.\cite{Heid1997} For instance,
experimental and theoretical studies of C$_{60}$ show that the lowest
energy internal vibrational mode energy $\approx$ 33~meV and
corresponds to the deformation of the Buckminsterfullerene sphere
into a prolate ellipsoid.\cite{HPark2000} Similar considerations have
been used to distinguish a known vibrational mode in
C$_{140}$.\cite{Pasupathy2005} In addition it is also possible to
excite a center-of-mass oscillation of a molecule (commonly known as
the bouncing ball mode) within the confinement potential that binds it
to the electrode surface\cite{HPark2000}.

\begin{figure}[!h]
\begin{center}
\includegraphics [clip, width=8cm]{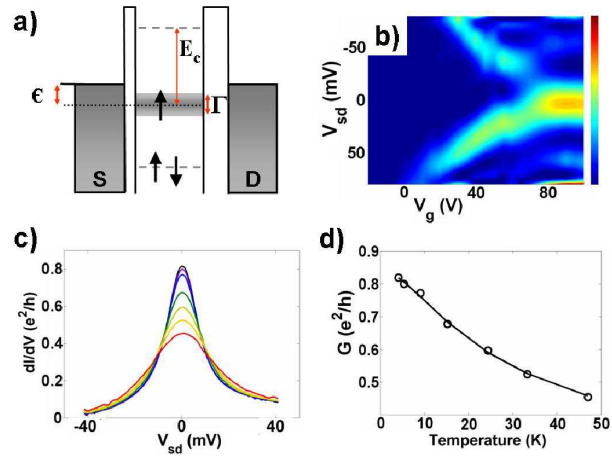}
\end{center}
\hspace\fill \vspace{-8mm} \caption{(a) Energy level diagram of a single molecule transistor device in the Kondo regime.  (b) Stability diagram of an SMT showing a charge state transition in which tunneling shifts from the Coulomb blockade regime to the Kondo regime.  Colorbar ranges from zero conductance (dark blue) to $\sim 0.5 e^2/h$.  (c) Traces of $dI/dV$ vs $V_{\mathrm{SD}}$ displaying a Kondo resonance at several different temperatures.  (d) Peak conductance of resonances in (c) as a function of $T$ fit to equation 4.  $T_{\mathrm{K}} \sim 58$~K.} \label{Kondo1}
\end{figure}

An additional higher-order tunneling process that is apparent in many
SMT devices is the formation of a many-body Kondo resonance.  This
phenomenon has studied in semiconductor quantum
dots\cite{Goldhaber1998,Cronenwett1998,Jeong2001} and
SMTs\cite{JPark2002,Liang2002,Yu2004b,Pasupathy2004,Yu2005,Natelson2006}.
Observing the Kondo effect requires an unpaired electron to exist in
the active element.  In the framework of the Anderson single level
impurity model, the active element is then an effective magnetic
impurity\cite{Kouwenhoven2001} with the singly occupied level's
energy, ${\epsilon}$, below the source/drain Fermi level (Figure
\ref{Kondo1}a).  Because of the Coulomb interaction, it is classically
forbidden to bring the electron out of the impurity without adding
energy into the system.  However, higher order tunneling processes can
take place such that another electron from the source/drain electrode
Fermi sea may exchange with the local moment.  At low temperatures the
coherent superposition of all possible cotunneling events results in
the screening of the local spin; the resulting ground state is a
correlated many-body singlet state spanning the source, impurity, and
drain.  This Kondo resonance is manifested as
a conductance peak at the Fermi energy ($V_{\mathrm{SD}}=0$).

The Kondo temperature, $T_{\mathrm{K}}$, is the characteristic temperature 
associated with the formation of the Kondo singlet.  
The characteristic energy scale $k_{\mathrm B}T_{\mathrm{K}}$ is
exponentially dependent on $\Gamma$,
 the intrinsic
width of the single particle level, by the relation\cite{Wingreen1994}
\begin{equation}
k_{\mathrm B}T_{\mathrm K} = \frac{\sqrt{{\Gamma}E_{\mathrm{c}}}}{2}e^{-{\pi}{\epsilon}(E_{\mathrm{C}} - \epsilon)/{\Gamma}E_{\mathrm{C}}}
\end{equation}
outside the mixed valence regime ($\epsilon\Gamma < 1$).  The total
width $\Gamma = {\Gamma}_{\mathrm{S}} + {\Gamma}_{\mathrm{D}}$ is
dictated by the overlap between the single particle state and the
conduction electron states of the source and drain.  Therefore
$\Gamma$ is in turn exponentially sensitive to the precise
molecule-electrode configuration.  This indicates both that a Kondo
resonance will only be found in devices with relatively strong
molecule-electrode coupling, and for SMTs displaying a Kondo
resonance, the overall conductance level will be significantly
increased with respect to similar SMTs in the Coulomb blockade regime.
In a symmetric system, as $T \rightarrow 0$, the SMT will approach its
theoretical maximum conductance $G_0 \equiv 2e^2/h$.  It is possible
that a relatively high Kondo temperature can be associated with a
device exhibiting a lower conductance as the total conductance is
determined by the smaller of ${\Gamma}_{\mathrm{S}}$ and
${\Gamma}_{\mathrm{D}}$ while $T_{\mathrm K}$ is determined by the
total $\Gamma$.

The Kondo resonance
will decrease in magnitude and increase in width as temperature is
increased.  A fit to the semiempirical expression for the
spin-$\frac{1}{2}$ Kondo resonance related to maximum conductance,
\begin{equation}
G(T) = \frac{G(0)}{(1 + (2^{1/s} - 1)(T/T_{\mathrm K})^2)^s},
\end{equation}
where $s = 0.22$, or to the full width at half max
\begin{equation}
{\mathrm{FWHM}} = {\frac{2}{e}}\sqrt{({\pi}{k_{\mathrm B}}{T})^2 + 2({k_{\mathrm B}}{T_{\mathrm K}})^2}.
\end{equation}
allows extraction of the corresponding Kondo temperature, $T_{\mathrm{K}}$.
Measurements of $G(T)$ or FWHM as a function of $T$ that cannot be fit to
these expressions likely indicate a zero-bias resonance that does not
originate from a Kondo state.  
Zeeman splitting of the zero-bias resonance in an applied
magnetic field can also be used to determine the magnetic nature of
molecular state, and hence to corroborate the presence of a Kondo
state.  Finally, we always attempt to verify that the measured
stability diagrams can be seen to transition from the Kondo regime to
the Coulomb blockade regime.  This evidence that charge state of the
device can be altered with the gate together with the data from the
tunneling thresholds of the potential SMT in the Coulomb regime will
further attest to the nature of the device.

The Kondo effect can be an additional ``fingerprint'' used to identify
molecular states.  Gate dependence of the Kondo resonance in SMTs has
proven to be weaker than that seen in semiconductor dots\cite{Yu2005},
and ``satellite'' peaks associated with Kondo physics in the presence
of molecular vibrations have been
observed\cite{Yu2004b,Yu2005,Parks2007}.  Also, since only odd
occupancy charge states can exhibit the Kondo effect, the presence of
a zero-bias resonance can be used to help identify specific charge
states.  Care must still be exercised, however: Zero-bias resonances
that do not result from the Kondo effect may occur in some devices,
while devices believed to contain metal nanoclusters\cite{Houck2005}
can exhibit the Kondo effect in the absence of molecules.

\section{Optical effects in electromigrated nanogaps}
\label{optical}
Combined optical and transport experiments on electromigrated nanogaps
can reveal a wealth of additional information beyond that available in
purely electronic measurements.  The same source/drain electrodes used
to couple current to the nanogap are observed to act as tremendously
effective plasmonic antennas, leading to dramatic surface enhanced
Raman scattering (SERS) in the junctions\cite{Ward2007,Ward2008}.
Here we describe recent results in combining SERS and transport
measurements, and report an additional effect: optical production of
dc electrical currents in these molecular nanogap systems.  These
currents are consistent with optical rectification due to nonlinearity
of the electromigrated junction's conduction, and provide a means of
estimating the magnitude of the enhanced optical fields in the
junction region.

\subsection{Surface-enhanced Raman scattering}
We have performed a series of optical measurements of Au nanogap
structures (prepared as in Section \ref{fab}) using a WITec CRM 200
Confocal Raman microscope.  Measurements were made with a 785~nm diode
laser with an incident power of $\approx$0.5~mW, chosen to maximize
signal, minimize photodamage to assembled molecules, and avoid
thermally driven rearrangement of the nanogap electrodes.  All optical
measurements were performed at room temperature in air.  Nanogap
devices were located in the microscope by rastering the sample beneath
the microscope objective to create a spatial map of the Raman response
with step sizes as small as 10~nm.  The Si substrate's strong
520~cm$^{-1}$ Raman peak can be used to map out the position of
nanogap since the Au film of the electrodes attenuates the Si Raman
response.  Once the nanogap was located, Raman spectra were taken with
1-2~s integration time periods to study the dynamics of the nanogap
system.

Initial experiments examined nanogaps as a potential surface-enhanced
Raman spectroscopy (SERS) substrate\cite{Ward2007}, with
para-mercaptoaniline (pMA) as the molecule of interest.
Nanoconstrictions were placed in parallel to allow simultaneous
electromigration of seven nanogaps at one time.  Samples were
characterized in the Raman microscope via spatial maps and time
spectra of the SERS response.  Prior to electromigration, no
significant SERS response is detected anywhere on the devices.

Following electromigration, we observe a SERS response strongly
localized to the resulting gaps.  Successive spectra measured directly
over the SERS hotspot revealed ``blinking'' and spectral diffusion,
phenomena often associated with single- or few-molecule Raman
sensitivity.  Blinking occurs when the Raman spectrum rapidly changes
on the second time scale with the amplitudes of different modes
changing independently of one another.  Spectral shifts as large as
$\pm20$~cm$^{-1}$ were observed, making it difficult to directly
compare SERS spectra with other published results.  Blinking and
spectral shifts are attributed to movement or rearrangement of the
molecule relative to the metallic substrate.  It is unlikely that an
ensemble of molecules would experience the same rearrangements
synchronously and thus blinking and wandering are expected to be
observed only in situations where a few molecules are probed.

More recently, we have performed simultaneous SERS and
transport measurements\cite{Ward2008}, including
Raman microscope observations over the center of nanogap
devices during electromigration.  Molecules of interest, pMA or a fluorinated
oligimer (FOPE), were assembled on the Au surface prior to
electromigration.  A 100$\times$ ultra-long working distance objective was
used to allow electrical probes to be placed beneath the objective to
make contact with the nanogap source and drain electrodes.  The nanogaps were
migrated \emph{in situ} using a computer controlled DAQ.
Transport
measurements were made by sourcing a 50-100~mV~RMS sine wave at 200~Hz
using a SRS~830 lock-in amplifer into one electrode, with a Keithley~428
current-to-voltage amplifier connected at the other electrode.
The ac current and its second harmonic were measured with
lock-in amplifiers and the dc current was measured at 5.0~kHz using a
DAQ.  Simultaneously Raman spectra were acquired with a 1~s integration
time.

Optical measurements during electromigration provide a wealth of
information about the plasmonic properties of nanogaps.  Once the
device resistance exceeds approximately 1~k$\Omega$, SERS can be seen.
This indicates that localized plasmon modes responsible for the large
SERS enhancements may now be excited.  As the gap further migrates the
SERS response is seen to scale logarithmically with the device
resistance until the resistance reaches approximately 1~M$\Omega$.  In
most samples the Raman response and conduction of the nanogap become
decoupled at this point with the conduction typically changing little
while uncorrelated Raman blinking occurs.  A more extensive
discussion of the connection between plasmonic modes and interelectrode
conductance is presented elsewhere\cite{Ward2008}.

\begin{figure}
\begin{center}
\includegraphics[clip,width=8cm]{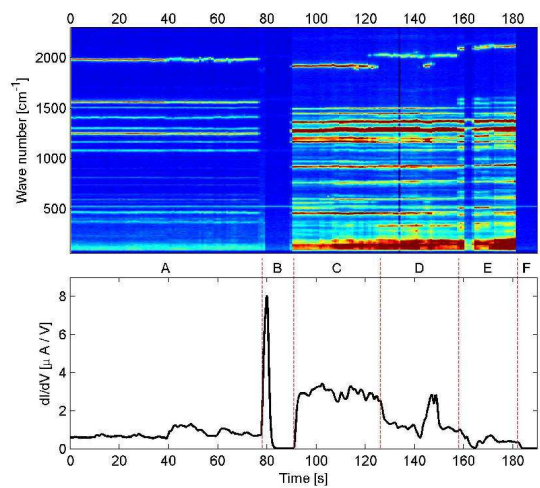}
\end{center}
\caption{\label{figWaterfall} Waterfall plot of Raman spectrum (1~s
  integration) and correlated conductance measurement for a single
  FOPE molecule shown at bottom.  The Raman mode observed between
  1950~cm$^{-1}$ to 2122~cm$^{-1}$ is believed to be for the same
  2122~cm$^{-1}$ mode associated with the C$\equiv$C stretch of the
  FOPE molecule.  The large spectral shifts observed for this mode are
  attributed to interactions between the molecule and its nanogap
  environment.  Clear correlations between the Raman structure and
  conductance can be seen.  In particular during region B and for part
  of region E the Raman spectrum is observed to disappear while the
  conductance drops to zero.  Between different regions distinct
  changes in spectrum are observed during clear changes in
  conductivity.  }
\end{figure}

In about 11\% of 190 devices, however, the Raman response and
conduction show very strong temporal correlations.  The 11\% yield is
is quantitatively consistent with yields of gateable SMTs as mentioned
in Section~\ref{transport}.  A typical correlated SERS time spectrum
and conductance measurement for a FOPE device are presented in figure
\ref{figWaterfall}.  The temporal correlations between SERS and
conduction are clear.  In region A we observe a stable Raman spectrum
and small conductance changes which appear to be correlated with
changes in the Raman mode at 1980~cm$^{-1}$.  After a short spike in
conductance the conductance and Raman disappear at B.  In section C
the conductance and Raman spectrum return.  This time the conductance
is 6x larger than in A and the Raman modes are different.  In
particular the mode previously seen at 1980~cm$^{-1}$ is now at
1933~cm$^{-1}$.  In region D we see the conductance drop to levels
closer to those seen at A and the 1933~cm$^{-1}$ mode from C shifts to
2038~cm$^{-1}$.  During D we see that the conductance momentarily
returns to the value seen in section C correlated with a shift in the
2038~cm$^{-1}$ mode back to 1933~cm$^{-1}$.  In section E we see
another shift in position of the 1933~cm$^{-1}$ mode to 2098~cm$^{-1}$
which slowly shifts up to 2122~cm$^{-1}$.  During E the Raman spectrum
is seem to once again disappear correlated with drop in conductivity
briefly and the return simultaneously.  Finally in F we see the
spectrum disappear a final time again correlated with a drop in
conductivity.

In the bulk Raman spectrum the FOPE molecule only shows one mode above
1700~cm$^{-1}$, the C$\equiv$C stretch mode at 2228~cm$^{-1}$
associated with the triple bond connecting the two
phenylene rings.  It is likely that
the mode at 2122~cm$^{-1}$ is a manifestation to the 2228~cm$^{-1}$
mode.  This mode frequently appears when
studying the FOPE devices and is absent in control experiments
with pMA or contaminants from the air.  The large shift in wave numbers
(over 100~cm$^{-1}$ between the bulk and normal and almost
300~cm$^{-1}$ for the greatest shift) indicates significant
interactions between the molecule and either the substrate
or other adsorbates, and is cause for further investigation.

Recalling that conduction in nanogaps is dominated by approximately a
single molecular volume, the observed correlations between conductance
and Raman measurements strongly indicate that the nanogaps have single
molecule Raman sensitivity.  It is then possible to confirm that
electronic transport is taking place through the molecule of interest,
via the characteristic Raman spectrum.  Data sets such as
Figure~\ref{figWaterfall} contain implicitly an enormous amount of
information about the configuration of the molecule in the junction,
and should be amenable to comparisons with theoretical calculations of
the optical properties of the molecule/electrode region.  These
combined optical and transport measurements open many possible
paths of exploration.

\subsection{Optically Induced Transport}
In addition to SERS, we also observe significant dc currents in
electromigrated nanogaps under illumination.
The mechanism of this optically induced transport may be probed
by studying the dependence of the dc current on the incident
optical intensity and the measured low frequency transport
properties of the junction.  
Resonant optical effects\cite{Galperin2005,Viljas2007} and photon-assisted tunneling\cite{Wu2006} are potential sources of dc optically-driven currents in
molecular junctions.  However, we find that the optically-driven dc
currents are relatively independent of molecule type, with similar
data sets collected in devices using, for example, pMA molecules, FOPE
molecules, and adsorbed atmospheric contaminants.  This strongly
suggests that the mechanism behind these optically-driven currents is
a general feature of the electromigrated junction structure, rather
than tied to specific molecular features.

One mechanism that is consistent with our observations is
\emph{optical rectification} due to the nonlinearity of the 
source/drain tunneling characteristics.  This effect has
long been considered in STM experiments\cite{Cutler1987}, though
its unambiguous observation has been very challenging\cite{Tu2006}.
The rectified 
current originates from the interaction of an ac excitation
with a nonlinear circuit element.  In the limit of a small bias $V$ the current 
can be approximated via a Taylor series.  In particular for 
an oscillating potential $V=V_0 + V_{\mathrm{ac}} \cos(\omega t)$:
\begin{equation}
I(V)=I(V_{0})+\left( \frac{\partial I}{\partial V} \right)_{V_{0}} V_{\mathrm{ac}} \cos(\omega t) + \frac{1}{2} \left( \frac{\partial ^2I}{\partial V^2} \right)_{V_{0}} {V_{\mathrm{ac}}}^2 \cos^2(\omega t) + \ldots
\label{eqnrect1}
\end{equation}
Applying a trigonometric identity,  
\begin{equation}
I(V)=I(V_{0})+\left( \frac{\partial I}{\partial V} \right)_{V_{0}} V_{\mathrm{ac}} \cos(\omega t) + \frac{1}{4} \left( \frac{\partial ^2I}{\partial V^2} \right)_{V_{0}} {V_{\mathrm{ac}}}^2 (\cos(2\omega t) + 1)+\ldots.
\label{eqnrect2}
\end{equation}
The conduction nonlinearity leads to a second-harmonic ac signal as
well as an additional dc current, both linearly proportional to
$\partial ^2I/\partial V^2$, which will depend on the device
geometry and conduction through the molecule.  Additionally the
optically rectified current will depend linearly on the incident laser
intensity.

Note that optical rectification in nanogap devices would allow an
experimental estimate of the enhanced optical field.  One 
can measure $\partial ^2I/\partial V^2$ using a low frequency 
(\emph{e.g.}, 200~Hz) ac signal.  If this nonlinearity results from
tunneling and the tunneling timescale is fast compared to an optical
cycle, then one can use the dc optically-driven dc current to 
infer $V_{\mathrm{opt}}$, the optical-frequency potential difference
across the source/drain electrodes at the point of tunneling.
With an estimate of the source/drain gap distance from $dI/dV$, an
estimate of the local plasmon-enhanced electric field is then
possible.

\begin{figure}
\begin{center}
\includegraphics[clip,width=8cm]{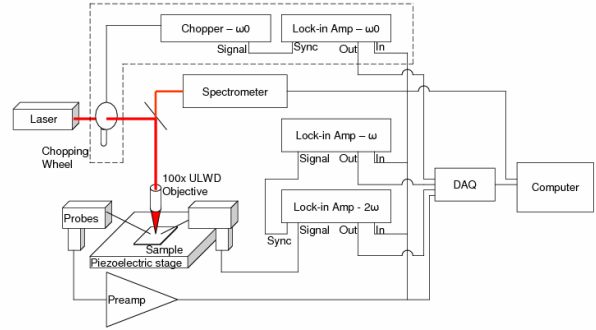}
\end{center}
\caption{\label{figSchematic} Schematic of optical/electronic
  measurement.  A 100~mV~RMS ac signal is applied by a lock-in into
  the source electrode at frequency $\omega$.  The ac current is
  measured at the drain by a current-to-voltage amplifier.  The dc
  current is measured from the amplifier by a DAQ at 5~kHz.  The ac
  current is measured at $\omega$ and $2\omega$ using lock-in
  amplifiers.  Raman spectra are acquired using 1~s integrations at an
  incident power of 0.5~mW at 785~nm.  For later experiments the
  equipment enclosed by the dashed lines was added to allow the laser
  to be chopped at frequency $\omega_0$.  $\omega$ and $\omega_0$ have
  been chosen such that they and $2\omega$ are at least 50~Hz apart.
  The optically rectified current is measured via a lock-in at the
  chopping frequency.  }
\end{figure}

Using the measurement scheme shown in Figure \ref{figSchematic} we
measure the first and second harmonics of the low-frequency current
using a lock-in amplifier, as well as the dc current.  In the absence
of any optical effect, (\ref{eqnrect2}) implies that the rectified
current due to the low-frequency drive will be exactly equal to the
lock-in current measured at the second harmonic of the source
frequency, as we observe.  Under illumination, plots of the dc current
\emph{vs.} the measured low-frequency ac current at $2\omega$ for
fixed input amplitude of 100~mV show that the dc current can exceed
the low frequency dc contribution by nearly a factor of two.  This
implies a second source of dc current that scales linearly with
$\partial ^2I/\partial V^2$ and vanishes when the 
illumination is blocked.

\begin{figure}
\begin{center}
\includegraphics[clip,width=8cm]{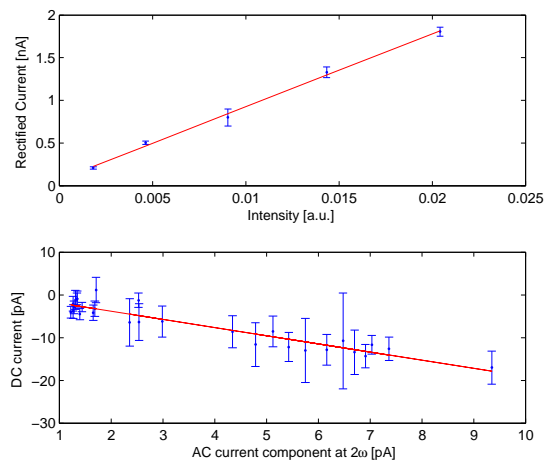}
\end{center}
\caption{\label{figOpticalPowerCurve} a) Rectified current as a
  function of optical intensity.  The measurement was performed on a
  nanogap assembled with pMA and migrated to a resistance of
  $\sim$10~M$\Omega$.  Error bars are one standard deviation from the
  mean value.  The data is well approximated with a linear fit
  consistent with the proposed model for optical rectification.  b) dc
  current measured as a function of the ac current component at
  2$\omega$ (source frequency).  Each point averaged over 2 seconds
  with the error bars indicating one standard deviation in the dc
  measurement over that time.  The data is reasonably represented by a
  linear fit with slope $-1.91\pm0.14$ with a r$^2$ value of 0.883.
  This slope magnitude in the presence of illumination implies
  significant optical rectification.  }
\end{figure}

To determine the particular optical mechanism at work, we measured the
dependence of the dc current on optical power.  We chop the incoming laser and
measure the current component at the chopping frequency as shown in
the Figure \ref{figSchematic}.  Figure \ref{figOpticalPowerCurve}
shows a power curve measured at a few intensities, showing that the
optically induced dc current depends linearly on the incident intensity,
consistent with the optical rectification mechanism.

Assuming that optical rectification is at work we can estimate the
enhancement of the electric field in the nanogap, as described above.
By comparing the measured dc current to the low frequency ac current
component at 2$\omega$ at constant laser power we can determine how
much of the rectified current is a result of optical rectification.
Figure \ref{figOpticalPowerCurve}B shows a representative curve.  The
dc current is well fit by a line of slope -1.91.  A slope of $\pm1.0$
would have indicated that all the dc current is due to rectification
of the applied low frequency signal.  The sign of the slope varies
from device to device depending on the sign of $d^2I/dV^2$ and thus
the direction of the rectified current.  It should be noted that both
the low frequency and optical frequency parts should be rectified in
the same direction.  A slope of -1.91 indicates that
$(V_{\mathrm{optical}}/V_0)^2 = 0.91$.  For this particular
  measurement $V_0$=100~mV yielding $V_{\mathrm{optical}}$=95~mV.
  Assuming a gap separation of 1~nm we can get the approximate optical
  field strength across the gap is $\approx1\times10^8$~V/m.  The
  incident unenhanced optical field is $\approx2\times10^5$~V/m,
  yielding a field enhancement of approximately 500, and a total Raman
  enhancement of roughly $6\times10^{10}$, consistent with the
  predications for the necessary enhancement to observe single
  molecule SERS.

Unfortunately at room temperature and under higher intensity laser
powers the nanogap stability is poor, making it difficult to perform a
more in-depth analysis.  Currently we do not have the capability to
perform low temperature optical measurements; however, it is possible
to examine these same rectification mechanisms at radio frequencies in
a low temperature probe station.  Optical rectification has been
successfully demonstrated at microwave frequencies in STM
measurements\cite{Tu2006}.  A measurement scheme analogous to that
present in Figure \ref{figSchematic} was used, with $\partial^2
I/\partial V^2$ found by a low frequency measurement and the RF
contribution to the rectified current measured by chopping the
incident RF.

\begin{figure}
\begin{center}
\includegraphics[clip,width=8cm]{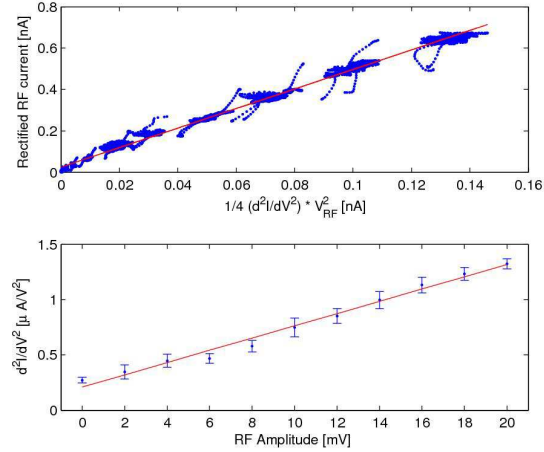}
\end{center}
\caption{\label{figRFMeasurements}
(a) RF Rectification vs $\frac{1}{4}\partial ^2I/\partial V^2 V_{\mathrm{RF}}$ 
measured at 10~MHz.  The rectified current scales linearly as expected by our 
model.  After factoring in the impedance mismatch between the RF function 
generator and the nanogap structure the data is fit well by a line of slope=1.  
This experimentally confirms equation \ref{eqnrect2} as the correct mechanism 
for rectification in nanogaps.  (b) $\partial ^2 I/\partial V^2$ as a 
function of the applied RF amplitude.  As the RF amplitude is increased the 
average nonlinearity increases in a linear fashion.
}
\end{figure}

Devices were prepared by assembling pMA on the Au surface prior to the
experiment.  Measurements were made at 80~K in a vacuum probe station
after electromigration.  A 20~mV~RMS low frequency sine wave was
sourced by one lock-in amplifier and fed into a bias tee along with a
0-20~mV~RMS RF signal.  The RF signal was amplitude modulated from 0
to 100\% a frequency $\omega_0$ (typically 157~Hz). The combined
signal was sent into one electrode of the nanogap.  The current flow
through the device was measured at the other electrode using a current
amplifier with 100~kHz low pass filter enabled to remove any RF
component from the measured current.  The output of the current
amplifier was measured using three lock-in amplifiers measuring at
$\omega$, $2\omega$, and $\omega_0$ (lock-in source frequency, second
harmonic, and RF chopping frequency) and the dc current was measured
directly at the amplifier.  A representative power curve plotting
$\frac{1}{4}(\partial ^2I/\partial V^2) V_{\mathrm{RF}}$ versus
$I_{\omega_0}$ is plotted in figure \ref{figRFMeasurements} for an RF
frequency of 10~MHz.  A clear linear dependence is observed.  The
sloped of 4.48 is a result of an impedance mismatch between the RF
function generator and the nanogap device.  After accounting for this
mismatch the slope is approximately 1 as expected for the
rectification mechanism.  From this series of measurements we infer
that $\partial ^2I/\partial V^2$ is frequency independent up to 1~GHz.
\itshape{A priori} \normalfont this does not imply that $\partial
^2I/\partial V^2$ is frequency independent up to optical frequencies,
though the results seen in figure \ref{figOpticalPowerCurve} are
consistent with this rectification mechanism beyond $\sim 10^{14}$~Hz.

\section{Conclusions and prospects}

Electromigrated gaps have proven to be an enormously useful tool in
probing electronic properties in single molecules, though the lack of
imaging techniques has meant that great care must be exercised in
interpreting transport data.  Pure electronic transport can contain
signatures that are distinct to the molecules being probed (charging
energies, vibrational resonances, Kondo physics).  The recent
observation that electromigrated junctions are highly effective
optical antennas has great potential.  Simultaneous measurements of
transport and Raman spectra in single molecules are now possible,
allowing the vibrational fingerprinting of the active electronic
element in the junction.  Detailed information about molecular
orientation and bonding within the junction may be inferrable from the
Raman data.  Optical measurements also bring additional electronic
transport mechanisms into play, as optical rectification measurements
demonstrate.  The promise of combined single-molecule electronic and
optical characterization suggests that electromigrated molecular
junctions have a bright future as tools for physics and physical
chemistry at the molecular scale.

\ack{DRW acknowledges support from the NSF-funded Integrative Graduate Research and Educational Training (IGERT) program in Nanophotonics.  GDS acknowledges support from the W M Keck Program in Quantum Materials.  NH and DN acknowledge support from the Robert A. Welch Foundation grants C-1220 and C-1636, respectively. DN also acknowledges NSF award DMR-0347253, the David and Lucille Packard Foundation and the Research Corporation.}

\end{document}